\documentstyle[preprint,psfig,aps]{revtex}
\tightenlines
\begin{document}
\title{$J/\psi$ suppression in heavy ion collisions --
interplay of hard and soft QCD processes}
\author{C.~Spieles$^1$\footnote{Supported by the 
Alexander v.~Humboldt Foundation}\footnote{Email: cspieles@lbl.gov },
R.~Vogt$^{1,2}$, L.~Gerland$^3$, S.A.~Bass$^4{}^*$, M.~Bleicher$^3$, 
L.~Frankfurt$^{3,5}{}^*$, M.~Strikman$^6$, H.~St\"ocker$^3$, W.~Greiner$^3$}

\address{$^1$~Nuclear Science Division,
Lawrence Berkeley National Laboratory,
Berkeley, CA 94720, USA}
\address{$^2$~Physics Department,
University of California at Davis, 
Davis, CA 95616, USA}
\address{$^3$~Institut f\"ur
Theoretische Physik,  J.~W.~Goethe-Universit\"at,
D-60054 Frankfurt a.M., Germany}
\address{$^4$~Department of Physics, Duke University,
Durham, N.C. 27708-0305, USA}
\address{$^5$~School of Physics and Astronomy, Tel Aviv University, 69978
Ramat Aviv, Tel Aviv, Israel}
\address{$^6$~Department of Physics, Pennsylvania State University,
University Park, PA 16802, USA}
\maketitle

\begin{abstract}
We study $J/\psi$ suppression in $AB$ collisions assuming that the
charmonium states evolve from small, color transparent configurations.
Their interaction with nucleons and nonequilibrated, 
secondary hadrons is simulated using the microscopic model UrQMD.
The Drell-Yan lepton pair yield and the $J/\psi/$Drell-Yan ratio
are calculated as a function of the neutral transverse energy 
in Pb+Pb collisions at 160~GeV and found to be in reasonable 
agreement with existing data.
\end{abstract}

\pagebreak

\section{Introduction}
Experimental data on the $AB$ and $E_T$ dependence of $J/\psi$-meson
production
from the CERN SPS
\cite{baglin,gonin,abreu,romana} 
exhibit tantalizing evidence for $J/\psi$ suppression when compared to hard
QCD production.
Do the $J/\psi$ data in Pb+Pb-collisions indicate the creation of a
deconfined phase of strongly interacting matter, {\it i.e.}  a quark-gluon
plasma (QGP) \cite{kharzeev,wong}?
Such QGP scenarios generally rely on the experimental observation of
deviations from the predictions of models of hadronic suppression.
Simple analytical models of $J/\psi$ suppression have used nuclear 
absorption alone
\cite{kharzeev} or absorption by nucleons and comoving secondaries
\cite{vogtplb98} to fit all but the Pb+Pb data. 
These models
generally assume a single, fixed nucleon absorption cross section for all charmonium
states and that the comover interaction rate, assuming a Bjorken scaling
expansion, is essentially thermal.
(For theoretical reviews, see
Refs.~\cite{reviewvogt,reviewkharzeev,reviewmuller}.)\footnote{Recently, there have been several attempts to build models of
charmonium production and absorption by means of microscopic 
hadronic transport simulations \cite{cassing,cassing2,geiss,kahana}. 
Based on rather different model treatments of the charmonium dynamics and the 
interconnection of hard and soft processes it is claimed in all these 
studies that conventional hadronic scenarios are consistent with the Pb+Pb
data.}


Charmonium final state interactions with mesons and
baryons are simulated using the Ultrarelativistic Quantum Molecular Dynamics, 
UrQMD \cite{bigpaper}, a microscopic hadronic transport model.
The charmonium nucleon cross sections employed here are calculated in a 
nonrelativistic potential model \cite{gerland} since
at SPS energies, charmonium-nucleon interactions are predominantly
nonperturbative. 
The charmonium precursors are not
eigenstates of the QCD Hamiltonian, therefore their effective sizes and
interaction cross sections may vary from their production to the final state
formation \cite{gerland}. 
Thus the charmonium absorption cross sections are assumed to
expand linearly with time until their asymptotic value is reached.
Since the produced particles in the collision have a non-thermal
distribution \cite{mycollspec}, comover interactions can rather effectively
dissociate the $J/\psi$. Because of abundant high mass resonances,
most of the meson induced $J/\psi$ dissociation processes are exothermic
\cite{mycollspec}. They account for an important fraction of the observed
suppression, larger in Pb+Pb collisions than the nuclear absorption
alone. 


\section{The model}

We apply perturbative QCD to the production of charmonium states by 
simulating nucleus-nucleus collisions in the impulse approximation.
The nuclear dependence of parton distribution functions is neglected.
The resulting 
space-time distribution of charmonium production points is inserted into the
evolving hadronic environment calculated with UrQMD \cite{mycollspec}
since the rare quarkonium
production processes are small perturbations on the heavy ion 
collision\footnote{Since we consider only exclusive charmonium production, 
the QCD factorization theorem is inapplicable.}.
Our model is thus designed to account for partonic and hadronic aspects of
the charmonium dynamics. 
The space-time distributions of hard and soft processes, {\it i.e.} in
particular charmonium production and absorption, 
may overlap.
Thus, in our model, a charmonium state produced in a hard process can be
dissociated by the interaction with a comoving hadron before 
this state leaves the nuclear environment and before all other hard
production processes are completed (in contrast to Ref.~\cite{kahana}).
The probability for such an event, however, is reduced due to the 
initially small absorption cross sections and the finite formation times of 
comoving mesons according to the string model (see below).
To avoid double counting, interactions of $c \bar c$ states with produced
hadrons in individual $NN$ collisions are excluded. In $AB$
collisions, all hard $NN$ collisions can contribute to the $c\bar c$
production while all soft $NN$ collisions can contribute to the hadronic
environment in which the $c \bar c$ state may be dissociated. The error
imposed by this concept --- inherent to all microscopic and analytical
models of comover absorption --- is estimated to be very small.

The charmonium states are distributed according to their assumed 
production probability 
times their decay probability to $J/\psi$'s.
Thus 40\% of the final states are $\chi$'s, 55\% are $J/\psi$'s, and 5\% are 
$\psi'$s
\cite{gavai}. According to the spin degeneracy, 
1/3 of the $\chi$'s are $\chi_{c10}$ states and 2/3 are $\chi_{c11}$ states.
Their momenta are assigned according to the parametrization \cite{ramona},
\[
E\frac{d\sigma}{dMdp^3}\sim(1-x_F)^{3.55} \exp(-p_T\, 2.08\rm \,GeV^{-1})\;
.
\]

The rescattering cross sections for $X(c\bar c)+B$, assuming $B\equiv N$,
are taken from Ref.~\cite{gerland}:
$\sigma(J/\psi N)=3.6$~mb, $\sigma(\psi' N)=20$~mb, 
$\sigma(\chi_{c10} N)=6.8$~mb, and
$\sigma(\chi_{c11} N)=15.
9
$~mb. Charmonium-meson cross sections 
($X(c \bar c) +\pi$, $X(c \bar c) +\rho$, {\it etc.})
are reduced by a factor of 2/3 from the corresponding baryon values.
All baryon and meson collisions above the respective dissociation 
thresholds are assumed to break up the charmonium state. 
Universal and energy independent cross sections are employed, ignoring any
charmonium-meson resonances, perhaps too 
crude an assumption. 
From phase space arguments one can infer that the $J/\psi$ dissociation cross
sections with $\pi$'s should be suppressed close to threshold while it
should be enhanced for exothermic channels. 
However, we have found that, during the initial stage of the $J/\psi$ comover
collisions, the average interaction energy, $<E>\approx 5$~GeV, is far 
above threshold \cite{mycollspec}.
There are no calculations of $J/\psi$ dissociation cross sections with
mesons other than $\pi$'s and $\rho$'s \cite{reviewmuller}.
In a thermal comover scenario the density of heavier mesons is suppressed 
by the Boltzmann factor. According to the UrQMD simulations, however,
these mesons dominate the comover absorption \cite{mycollspec}.
In a forthcoming paper~\cite{paper} the influence of the energy 
dependence of the comover interactions will be studied further.

The cross sections correspond to the geometrical transverse radii
$r_T^i=\sqrt{\frac{\sigma^i}{\pi}}$ of the
charmonium states. 
We use $\sigma^i$  to estimate the respective 
formation times $\tau_F^i$ of the charmonium states by choosing
$\tau_F^i=r_T^i/c$.  
During these formation times 
the cross sections increase linearly with $t$ \cite{gerland}, 
starting from 0 at $t=0$.

Here it is important that we also take into account the formation 
time of comoving mesons (on average, $\tau_F\approx 1$ fm/c). Particles 
produced by string fragmentation are not allowed to interact with other hadrons --
in particular with a charmonium state -- within their formation time. However,
leading hadrons are allowed to interact with a reduced cross section even
within their formation time. The reduction factor is 1/2 for mesons which
contain a leading constituent quark from an incident nucleon and 2/3 for
baryons which contain a leading diquark.

For this study, we have slightly modified the angular distributions of 
meson-baryon interaction in the UrQMD~1.0 model since their strong forward peak
underpredicts the total produced transverse energy \cite{bleicher}.
The model now reproduces the $E_T$ spectra in 
S(200~GeV)+Au and Pb(160~GeV)+Pb collisions measured by NA35 and NA49,
respectively \cite{unpubl}.
Neither the amount of baryon stopping nor the rapidity distribution of 
negatively charged particles 
which have been shown to agree with experimental $pp$ and $AB$ interactions
\cite{bigpaper} are significantly affected by this 
modification. 


\section{Results and discussion}

\begin{figure}[b]
\vspace*{\fill}
\centerline{\psfig{figure=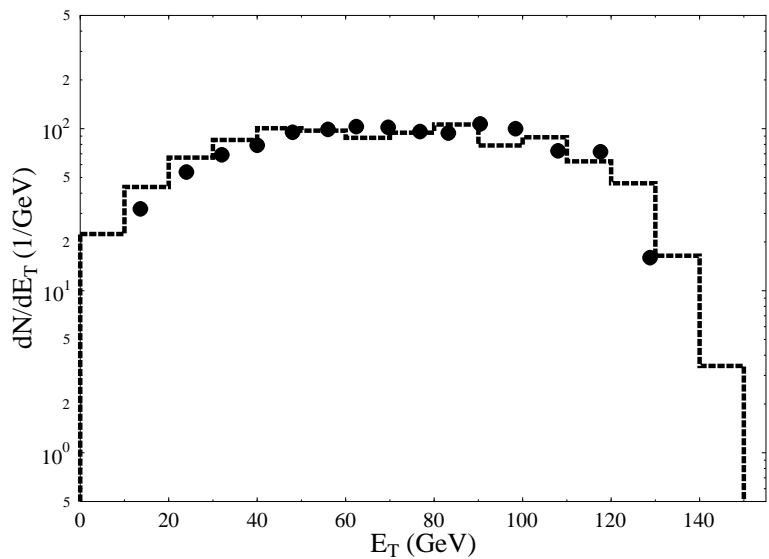,width=12cm}}
\caption{
Number of Drell-Yan pairs in Pb+Pb interactions as a function of the
 neutral transverse energy within $1.1<\eta<2.3$. 
The calculation is normalized to the data.
Shown is the UrQMD result and experimental data from NA50
\protect\cite{ramello} with the
$E_T$ of the data rescaled by $0.8$. The modification is motivated by 
a comparison of the
recently published $E_T-E_{ZDC}$ contour plot \protect\cite{romana}
with the previously published analysis
\protect\cite{abreu,ramello}.
\label{hardet}}
\vspace*{\fill}
\end{figure}

Figure~\ref{hardet} shows the calculated number of Drell-Yan muon pairs, 
proportional to the number of hard collisions, in Pb+Pb collisions
as a function of the produced neutral transverse energy within
$1.1<\eta<2.3$. The NA50 data
\cite{ramello} have been included
with the abscissa rescaled to reflect the latest change in
data \cite{romana} which indicates an $\approx 20$~\% shift in the 
absolute $E_T$ scale from previous publications \cite{abreu,ramello}.
We are aware that the new analysis does not imply a simple
overall rescaling of the old data points. However, in order to reasonably
compare the gross features of the
experimental dimuon $E_T$ spectrum  
with our model calculation, we have multiplied all $E_T$-values of the data
by 0.8.
This factor was obtained by comparing the $E_T-E_{ZDC}$ contour from
Quark Matter '97 \cite{ramello} and the Moriond '98 \cite{romana} proceedings.
The  agreement between the model and the rescaled data
is to be expected since the NA49 $E_T$ distribution \cite{na49} is 
described correctly \cite{unpubl}\footnote{However, the agreement between the 
model and 
the NA38 experiment becomes poor for S+U collisions. 
The UrQMD calculation appears
to overestimate the neutral transverse energy by about 25\%
in the range $1.7<\eta<4.1$ \cite{baglin,borhani}
although the calculated $E_T$ spectrum of the similar
S+Au system agrees well with the NA35 data \cite{na49}. 
The UrQMD calculation thus indicates an inconsistency between the $E_T$
measurements by NA35, NA49 and NA50 on the one hand and NA38 on the other
\cite{unpubl}.}. The additional 
simulation
is a simple and well understood model of hard scattering 
processes in nucleus-nucleus collisions.

\begin{figure}[b]
\vspace*{\fill}
\centerline{\psfig{figure=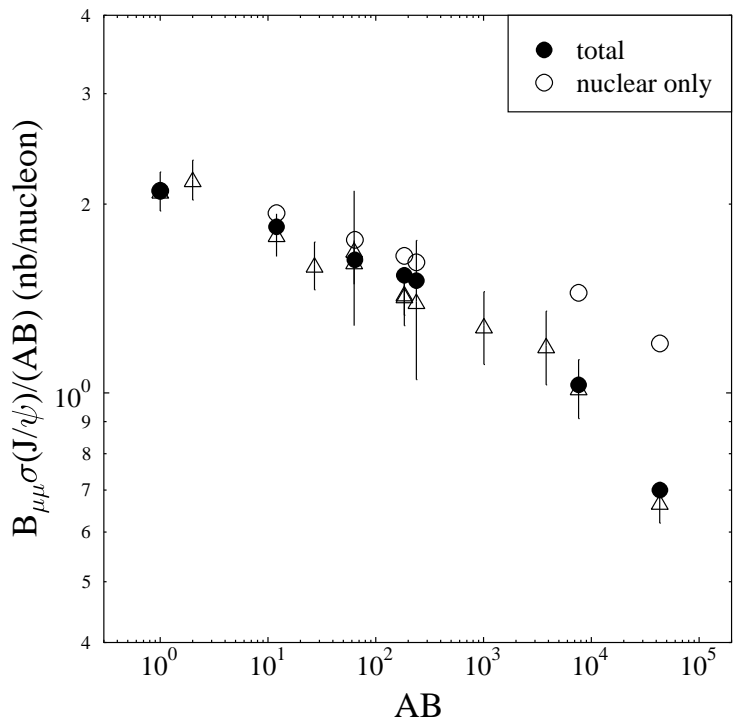,width=12cm}}
\caption{$J/\psi$-production cross sections times dimuon branching ratio
 in the
kinematical domain $0<y_{cm}<1$ and $|\cos\theta_{CS}|<0.5$, and rescaled,
if necessary, to $p_{\rm lab}=200$~GeV as a function of
$AB$. The data (open triangles) are from \protect\cite{abreu}.
Open circles denote the production cross
sections if only nuclear absorption is considered.
\label{sigab}}
\vspace*{\fill}
\end{figure}

Figure~\ref{sigab} shows the $J/\psi$ production cross section 
according to 
UrQMD calculations for several projectile-target combinations 
($p$(450~GeV)+C, $p$(200~GeV)+Cu, $p$(200~GeV)+W, $p$(200~GeV)+U, 
S(200~GeV)+U, and Pb(160~GeV)+Pb) in comparison to experimental data
\cite{abreu}.
The results of the calculations are normalized to the experimental
cross section in $p$(200~GeV)+$p$ interactions. The 450~GeV and 160~GeV 
simulations are rescaled to $p_{lab}=200$~GeV with the parametrization of 
Ref.~\cite{schuler}, as done by NA50 \cite{gonin,abreu}. 
Considering only nuclear dissociation results in a far smaller $J/\psi$ 
suppression than seen in the data, not only for Pb+Pb collisions but also
for S+U and even $pA$ reactions. Note that the systematics of
nuclear absorption shown in Fig.~\ref{sigab} does not reflect a universal
straight line as in Glauber calculations with constant absorption cross
sections.
By taking the nonequilibrium charmonium-meson interactions into
account, good agreement with the data is obtained. However, a strong 
dependence on parameters such as the charmonium and comover formation 
times and the dissociation cross sections remains to be 
studied in detail \cite{paper}.

Due to the linear expansion of the charmonium cross sections with 
time, the $J/\psi$ and $\psi'$ cross sections are similar in the very
early stages, leading to a weak $A$-dependence of the $\psi'/J/\psi$
ratio in $pA$ collisions 
at central rapidities, apparently consistent with the data.
However, for a deeper understanding of this ratio quantum interference
effects \cite{lonya} as well as refeeding processes, $\pi J/\psi \rightarrow
\psi' \pi$ \cite{sorge,chen}, must be considered.
The $\psi'/J/\psi$ ratio will be studied further later \cite{paper}.

\begin{figure}[b]
\vspace*{\fill}
\centerline{\psfig{figure=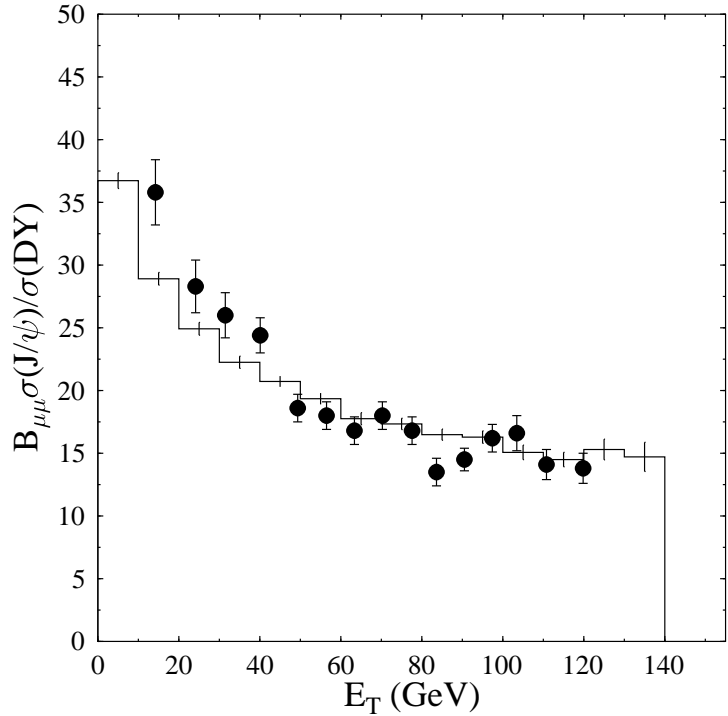,width=12cm}}
\caption{The ratio of $J/\psi$ to Drell-Yan production as a function of 
$E_T$ for Pb+Pb at 160~GeV. 
The experimental data are from Ref.~\protect\cite{romana}.
The normalization factor, from $pp$ interactions at 200~GeV, $B_{\mu\mu}\sigma(
J/\psi)/\sigma({\rm DY})=46$ 
is taken from Ref.~\protect\cite{kharzeev}. 
This value, however, has been indirectly determined in the framework 
of a different model. An additional factor of 1.25 
\protect\cite{reviewvogt} has been applied to the Pb+Pb calculation
in order to account for the lower energy. Note that no scaling factor has
been applied to the $x$-axis for either the calculations or the data.
\label{psidyet}}
\vspace*{\fill}
\end{figure}

Figure~\ref{psidyet} shows the $J/\psi$ to Drell-Yan ratio 
as a function of $E_T$ for Pb+Pb interactions at 160~GeV compared to 
the NA50 data \protect\cite{romana}.
The normalization of $B_{\mu\mu}\sigma(J/\psi)/\sigma({\rm DY})=46$ in $pp$
interactions at 200~GeV has been 
fit  to S+U data within a geometrical
model \protect\cite{kharzeev}.
The application of this value to our analysis is not arbitrary:
the model of Ref.~\protect\cite{kharzeev} renders 
the identical $E_T$-integrated $J/\psi$ survival probability, $S=0.49$, 
as the UrQMD calculation for this system.
An additional factor of 1.25
\protect\cite{reviewvogt} has been applied to the Pb+Pb calculation
in order to account for the lower energy, 160 GeV, since the
$J/\psi$ and Drell-Yan cross sections have different energy and isospin
dependencies. 
The gross features of the $E_T$ dependence of the $J/\psi$ to Drell-Yan
ratio are reasonably well described by the model calculation. 
No discontinuities in the shape of the ratio as a function of $E_T$ 
are predicted by the
simulation. 

\section{Conclusion}

We have examined charmonium production and absorption processes in
$pA$ and $AB$ collisions at SPS energies.
The microscopic simulation of hard processes in the impulse 
approximation and the hadronic transport description of 
$AB$ collisions with the UrQMD model
simultaneously provide reasonable $E_T$
dependencies of the Drell-Yan rates as well as
baryon and meson rapidity distributions. We have modelled $J/\psi$
absorption according to the scenario described in Ref.~\cite{gerland}.
Cross sections evolving from color transparent small configurations to
asymptotic states derived from quantum diffusion and the 
dynamical $\chi$ polarization (color filtering) are taken into account.

The calculated $J/\psi$ production cross sections for minimum bias $pA$,
S+U and Pb+Pb collisions agree with experiment. Dissociation by
nonequilibrium 
comovers accounts for about half of the total absorption in 
S+U and Pb+Pb reactions.
The contribution of the interaction with comovers
in $pA$ reactions is small but not negligible.
The suppression of charmonium states is sensitive 
to the comover momentum distributions. The effective dissociation by comovers
seems to indicate a nonequilibrated hadronic environment.
The observed $E_T$ dependence of the $J/\psi$ to Drell-Yan ratio in Pb+Pb
collisions is reproduced by the model. The calculated result is
smooth, without abrupt discontinuities,
in agreement with new high statistics data~\cite{kluberg}.
We conclude that within our model, the data on charmonium cross 
sections at the SPS can be explained without invoking exotic mechanisms.

\end{document}